\documentclass[reprint,aps,amssymb]{revtex4-2}
\usepackage{graphicx}
\usepackage{color}
\usepackage{mathrsfs}
\usepackage{multirow}
\usepackage{dcolumn}
\usepackage{bm}
\usepackage{amsmath}
\usepackage[svgnames]{xcolor} 
\usepackage[colorlinks=True, linkcolor=SlateBlue,
            citecolor=SteelBlue,urlcolor=BlueViolet]{hyperref}
\usepackage[capitalise]{cleveref}

\newcommand{\rms}{\rm s}

\newcommand{\mnras}{Mon. Not. R. Astron. Soc.}

\newcommand{\aap}{Astron. Astrophys.}
\newcommand{\baas}{B. Amer. Astron. Soc.}
\begin{document}

\title{Probing the Nature of Dark Matter via Gravitational Waves Lensed by Small Dark Matter halos}

\author{Xiao Guo} 
\affiliation{National Astronomical Observatories, Chinese Academy of Sciences, 20A Datun Road, Beijing 100101, China}
\affiliation{School of Astronomy and Space Science, University of Chinese Academy of Sciences, 19A Yuquan Road, Beijing 100049, China}

\author{Youjun Lu}
\email{luyj@nao.cas.cn}
\affiliation{National Astronomical Observatories, Chinese Academy of Sciences, 20A Datun Road, Beijing 100101, China}
\affiliation{School of Astronomy and Space Science, University of Chinese Academy of Sciences, 19A Yuquan Road, Beijing 100049, China}

\date{\today}

\begin{abstract}
Dark matter (DM) occupies the majority of matter content in the universe and is probably cold (CDM). However, modifications to the standard CDM model may be required by the small-scale observations, and DM may be self-interacting (SIDM) or warm (WDM). Here we show that the diffractive lensing of gravitational waves (GWs) from binary black hole mergers by small halos ($\sim10^3-10^6M_\odot$; mini-halos) may serve as a clean probe to the nature of DM, free from the contamination of baryonic processes in the DM studies based on dwarf/satellite galaxies. The expected lensed GW signals and event rates resulting from CDM, WDM, and SIDM models are significantly different from each other, because of the differences in halo density profiles and abundances predicted by these models. We estimate the detection rates of such lensed GW events for a number of current and future GW detectors, such as the Laser Interferometer Gravitational Observatories (LIGO), the Einstein Telescope (ET), the Cosmic Explorer (CE), Gravitational-wave Lunar Observatory for Cosmology (GLOC), the Deci-Hertz Interferometer Gravitational Wave Observatory (DECIGO), and the Big Bang Observer (BBO). We find that GLOC may detect one such events per year assuming the CDM model, DECIGO (BBO) may detect more than several (hundreds of) such events per year, by assuming the CDM, WDM (with mass $>30$\,keV) or SIDM model, suggesting that the DM nature may be strongly constrained by DECIGO and BBO via the detection of diffractive lensed GW events by mini-halos. Other GW detectors are unlikely to detect a significant number of such events within a limited observational time period. However, if the inner slope of the mini-halo density profile is sufficiently steeper than the Navarro-Frenk-White (NFW) profile, e.g., the pseudo-Jaffe profile, one may be able to detect one to more than hundred such GW events by ET and CE.  
\end{abstract}

\maketitle

\section{Introduction}

The cold dark matter (CDM) model can successfully reproduce the observed large scale structure, but has difficulties in interpreting small scale structures \cite{2018PhR...730....1T}. Searches of CDM particles by ground-based experiments have also excluded a large parameter space for weakly interacting massive particles \cite{2017NatPh..13..212L}, perhaps the most promising CDM candidates \cite{Bahcalletal2004}. These urge investigations on alternative dark matter (DM) models, such as self-interacting CDM (SIDM) \cite{2000PhRvL..84.3760S}, warm DM (WDM) \cite{2001ApJ...556...93B}, or fuzzy/wave-like DM \cite{2017PhRvD..95d3541H}, proposed to generate halos with core-like density profiles and/or smaller abundance at low-masses relative to the CDM model. 

The abundance and density profile of low-mass halos can be inferred via observations of dwarf/satellite galaxies \cite{2020arXiv201108865N,*2014MNRAS.442.2487K}. However, it may be biased due to the faintness of these galaxies and contamination from not-well understood baryonic processes \cite{1996MNRAS.283L..72N,2019MNRAS.488.2387B}. Halos with mass $\sim10^3-10^6M_\odot$ (hereafter mini-haloes), dark and free from complex baryonic processes, are ideal systems to study DM \cite{2014ApJ...792...99S,*2016MNRAS.456...85S}, but hard to observe by electromagnetic (EM) waves, even using its gravitational lensing effect \cite{1995ApJ...442...67U}.

Recently, it was shown that the lensing of gravitational waves (GWs) is an unique method to probe DM halos because the lensing effect leads to detectable waveform changes and resolvable time delays \cite{2018PhRvD..98j4029D, 2020ApJ...901...58O, 2022arXiv220400814O, 2022MNRAS.512....1G, 2021PhRvD.104f3001C, 2022MNRAS.509.1358U}. \citet{2021MNRAS.502L..16C, 2022A&A...659L...5C} also proposed that the strong lensing of GWs can be used to probe fluid DM. Not only a significant number of GW events strongly lensed by galaxies are expected to be detected by future GW detectors \cite{2018MNRAS.476.2220L,2014JCAP...10..080B}, but also GW events diffractively lensed by mini-haloes may be detectable \cite{2018PhRvD..98j4029D}. Here we show that the gravitational lensing of GWs by mini-haloes may be used to reveal the nature of DM. We investigate the lensed GW signals caused by mini-halos with different density profiles and estimate the event rates of such phenomena by assuming different DM models. The significant differences in the lensed GW signals and event rates resulting from different models clearly demonstrate that the gravitational lensing of GWs can serve as a clean probe to DM nature.

This paper is organized as follows. In Section~\ref{sec:signal}, we illustrate the diffractive lensing effects on GW signals by mini-halos with different mass density profiles. In section~\ref{sec:SNR}, we introduce the method for identifying the lensing signatures by mini-halos, using the SNR difference of the lensed signals from the unlensed ones. In Section~\ref{sec:rate}, the detectable lensing rates for stellar binary black hole (sBBH) merger events are estimated by assuming different DM models for the current and future GW detectors. The conclusions are summarized in Section~\ref{sec:sum}. 

\section{GW signal lensed by halos}
\label{sec:signal}

Consider a lens system in the wave optics regime, the source (GW event) and lens (DM halo) are locating at redshift $z_{\rms}$ and $z_{\ell}$, correspondingly distance $D_{\rms}$ and $D_{\ell}$ (lens to source $D_{\ell\rms}$), respectively. The lensed GW signal is 
\begin{equation}
\tilde{\phi}^L(f)=F(f)\tilde{\phi}(f)
\end{equation}
in the frequency domain, where $F(f)$ denotes the factor of (original) unlensed signal $\tilde{\phi}(f)$ amplified by the lens potential. In general, we have
\begin{equation}
F(w,\boldsymbol{y})=\frac{w}{2\pi i}\int d^2\boldsymbol{x}\exp[iw T(\boldsymbol{x},\boldsymbol{y})],
\label{eq:AmpFact}
\end{equation}
where $w$ is a dimensionless frequency (defined later for lenses with different density profiles), $T(\boldsymbol{x},\boldsymbol{y})=\frac{1}{2}|\boldsymbol{y}-\boldsymbol{x}|^2+\psi(\boldsymbol{x})-T_1(\boldsymbol{y})$, $\boldsymbol{x}=\frac{\boldsymbol{\xi}}{\xi_0}$ and $\boldsymbol{y}=\frac{D_{\ell}}{\xi_{0} D_{\rms}} \boldsymbol{\eta}$ are dimensionless angular coordinates on the lens and source planes, respectively, $\boldsymbol{\xi}$ the impact parameter, $\boldsymbol{\eta}$ the source position vector, $\xi_0$ a normalization length, $w$ the dimensionless frequency (defined below for different density profiles), $\psi(\boldsymbol{x})$ the lens potential (determined by the mass density profile of the lens) \cite{2001astro.ph..2341K}, $T_1(\boldsymbol{y})$ the arrival time of first image under geometrical optics. Obviously diffractive lensing can cause significant frequency-dependent amplification and phase modulation of waveforms  \cite{2019PhRvL.122d1103J, 2017arXiv170204724D, 2019PhRvL.122d1103J, 2020PhRvD.101f4011H, 2020PhRvD.101l3512D}. The calculation of amplification factor can refer to \cite{Guo2020}.

When $w\rightarrow\infty$, i.e., in the geometrical limit, only stationary points of Fermat potential (or time delay surface $T(\boldsymbol{x},\boldsymbol{y})$) contribute to the diffraction integral (Eq.~\ref{eq:AmpFact}). These stationary points satisfy $\nabla_xT(\boldsymbol{x},\boldsymbol{y})=0$, i.e., the lens equation
$
\boldsymbol{y}=\boldsymbol{x}-\nabla_x\psi(\boldsymbol{x}).
$
Thus the amplification factor is the summation of multiple images, 
\begin{equation}
F_{\rm geo}(w, \boldsymbol{y})=\sum_{j}\sqrt{|\mu_{j}|} \exp \left[i w T_{j}-i \pi n_{j}\right],
\end{equation}
where $T_j=T(\boldsymbol{x}_j,\boldsymbol{y})$ and $\mu_j=\mu(x_j)$ are for the position and magnification of the $j$-th image $\boldsymbol{x_j}$, $n_j=0$, $1/2$, and $1$ for the image position at the minimum, saddle, and maximum points of the Fermat potential, respectively (see \cite{2017arXiv170204724D,1999PThPS.133..137N}).

For the spherical symmetrical cases, the lens equation can be reduced to $y=x-d\psi(x)/dx$. By setting $dy(x)/dx=0$, i.e., $d^2\psi(x)/dx^2=1$, we obtain the critical value $x_{\rm cr}$, thus correspondingly $y_{\rm cr,g}$, which represents the boundary between cases with single image and double images.

\subsection{The pseudo-Jaffe lens}

In the strong lens studies, the pseudo-Jaffe profile is frequently adopted \cite{2001astro.ph..2341K}, i.e., 
\begin{equation}
\hat{\rho}_{\rm pJ}(x)= (x^2+s^2)^{-1}(x^2+a^2)^{-1},
\end{equation}
where $s$ and $a$ ($>s$) represent core and transition radius, respectively \cite{1983MNRAS.202..995J}. Though small halos may not follow the pseudo-Jaffe  profile, we still adopt such a profile for analysis below as reference. 

We define $\rho_{\rm pJ}(x)=\rho_{\rm s}\hat{\rho}_{\rm pJ}(x)$, where $\rho_{\rm s}$ is the scale density and $\rho_{\rm pJ}(1)=\frac{\rho_{\rm s}}{(1+s^2)(1+a^2)}$. For convenience, we adopt the scale radius as the same as the Einstein radius of a corresponding singular isothermal sphere (SIS) model with the same $\rho_{\rm s}$ but with $s\rightarrow0$ and consider the approximation when $s\ll x\ll a$ (or $a\rightarrow\infty$ but we still keep $a$ in the expression), i.e.,
\begin{equation}
\xi_0=4\pi\frac{\sigma_{\rm v}^2}{c^2}D_{\rm eff} = 4\pi\frac{\sigma_{\rm v}^2}{c^2} \frac{D_{\ell}D_{\ell\rm s}}{D_{\rm s}}.
\label{eq:xi0_pJ}
\end{equation}
This scale radius is adopted for convenience though it is not the real Einstein radius of the pseudo-Jaffe lens. The Einstein radius of a pseudo-Jaffe lens can be defined as $y_{\rm cr,g}\xi_0$, e.g., $y_{\rm cr,g}\simeq 0.168$ for $(s, a)=(0.1, 2)$; $y_{\rm cr,g}\simeq 0.637$ for $(s, a)=(0, 2)$. 

By analogizing the SIS model, we can substitute  $\sigma_{\rm v}$ in Equation~\eqref{eq:xi0_pJ} with $\rho_{\rm s}$, thus
$$
\xi_0=\frac{a^2c^2}{(4\pi)^2GD_{\rm eff}\rho_{\rm s}}
$$ 
since the enclosed mass within radius $x$ is $M(x)=\frac{\sigma_{\rm v}^2\xi_0 x}{G}$ for the SIS model. Then the total mass of the pseudo-Jaffe lens and the scale radius are related to each other by $M_\ell= \frac{2\pi^2 \rho_{\rm s} \xi_0^3}{a+s}$. Finally, we obtain 
\begin{equation}
\xi_0^2=\frac{8(a+s)GM_\ell D_{\rm eff}}{a^2c^2}.
\label{eq:xi0_pJ_M_l}
\end{equation}
Therefore, the dimensionless frequency and the time delay in Equation~\eqref{eq:AmpFact} are 
$$
w=\frac{2\pi f(1+z_\ell)}{cD_{\rm eff}}\xi_0^2=\frac{16\pi f(1+z_\ell)(a+s)GM_\ell }{c^3a^2},
$$
and
$$
T=\frac{cD_{\rm eff}t_{\rm d}}{\xi_0^2}=\frac{a^2c^3t_{\rm d}}{8(a+s)GM_\ell},
$$
where $t_{\rm d}$ is the real time delay. The dimensionless mass density for the pesudo-Jaffe profile is 
$$
\kappa(x)=\frac{\kappa_s}{2}\left[\frac{1}{\sqrt{s^{2}+x^{2}}}-\frac{1}{\sqrt{a^{2}+x^{2}}}\right],
$$
where $\kappa_s=\frac{a^2}{2\pi}$.
The lens potential can be expressed analytically as
$$
\begin{aligned}
\psi(x)=&\kappa_s\left[(\sqrt{s^2+x^2}-s)-(\sqrt{a^2+x^2}-a)\right.\\
-&s\ln\frac{s+\sqrt{s^2+x^2}}{2s} +\left. a\ln\frac{a+\sqrt{a^2+x^2}}{2a}\right].
\end{aligned}
$$

\begin{figure*}
\includegraphics[width=0.9\textwidth]{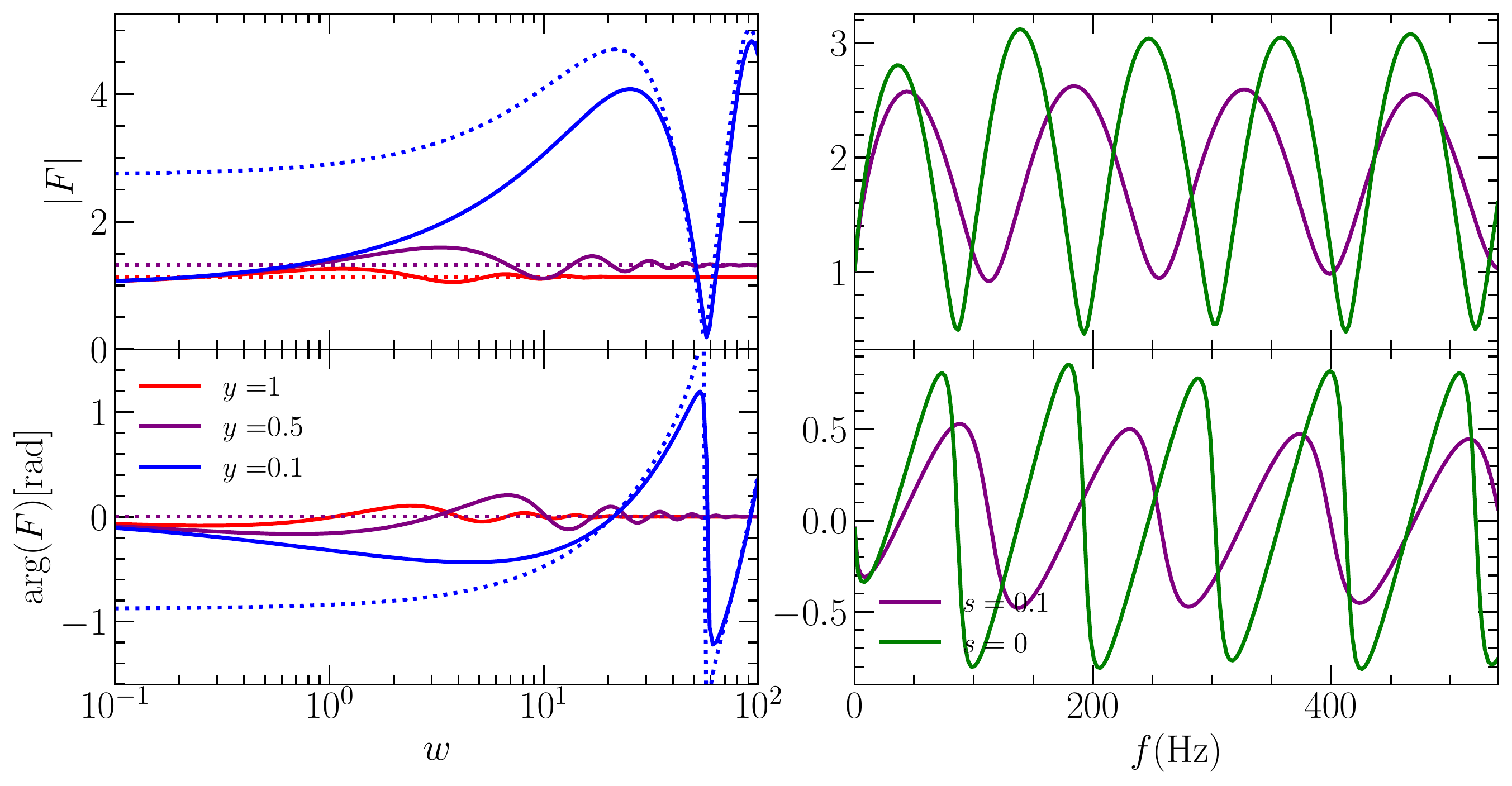} 
\caption{Magnitude (top panels) and phase (bottom panels) of the amplification factor obtained for pseudo-Jaffe lenses at $z_{\ell}=1$. Left panels: $(s, a)=(0.1, 2)$, for which $y_{\rm cr,g}\approx0.168$. Right panels: $(s, a)=(0.1, 2)$ or $(0, 2)$, $y=0.2$, $M_\ell=10^3M_\odot$, and $\xi_0=0.351$\,pc (for $s=0$, $y_{\rm cr,g}\approx0.637$). Solid and dotted lines represent the results obtained from wave optics and geometrical optics approximation, respectively.
} 
\label{fig:F_pJ} 
\end{figure*}

\begin{figure}
\includegraphics[width=0.45\textwidth]{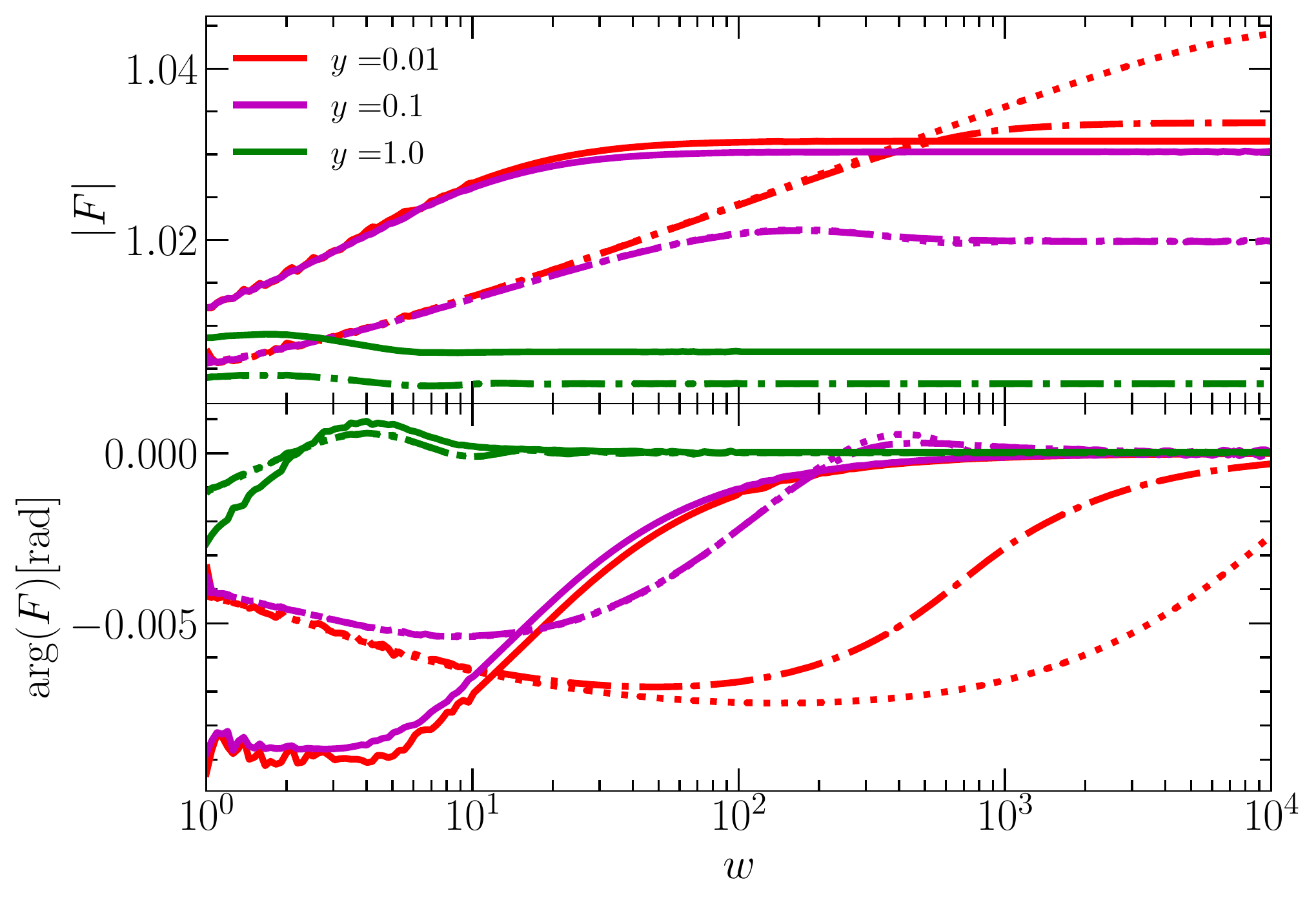} 
\caption{Magnitude (top panel) and phase (bottom panel) of the amplification factor for the NFW lens (dotted lines) and the IC-NFW lens with $s=0.1$ (dotted-dashed line) and $s=1$ (solid line) at $z_\ell=1$ with different $y$-values. The masses of all lenses are set as $M_\ell=10^3M_\odot$, for which $y=1/2$ corresponds to 1.27\,pc ($r_{\rms}$). The lensed waveforms by the IC-NFW lens, especially at low frequency ($w<300$), is similar to that by the NFW lens with the same mass when the size of core $s$ is small (e.g., $s=0.1$), however, it can be significantly different from that by the NFW lens when $s$ is large (e.g., $s=1$). This suggests that these two different density profiles may be distinguishable from each other by the difference in the lensed GW signals.
}
\label{fig:F_NFW_sCDM} 
\end{figure}

\subsection{The NFW lens and the CDM model}

In the CDM model, those DM  halos with mass $M_{\ell}$ may be actually described by the Navarro-Frenk-White (NFW) profile as 
\begin{equation}
\label{eq:NFW}
\rho_{\rm NFW}(x)=\rho_{\rms} x^{-1}\left(1+x\right)^{-2},
\end{equation}
where $\rho_{\rm s}$ is the scale density, $x=r/\xi_0$, $\xi_0=r_{\rm s}$ the scale radius for NFW profile \cite{NFW1996}, determined by the halo mass and concentration parameter $c_{\rm v}$ ($ = r_{\rm vir}/r_{\rm s}$ and $r_{\rm vir} $ is the virial radius of the halo) as given in  \cite{2001ApJ...559..572O} (see also \cite{Wang2019}), i.e., 
\begin{equation}
c_{\rm v}\left(M_{\ell}, z_\ell\right)=\frac{8}{1+z_\ell}\left(\frac{M_{\ell}}{10^{14} h_0^{-1} M_{\odot}}\right)^{-0.13}.
\label{eq:c_vLi}
\end{equation}
Here $h_0=0.677$ is dimensionless Hubble constant. Adopting the NFW profile, the lens potential is $\psi(x)=\kappa_{\rm s}\hat{\psi}(x)$, where $\kappa_{\rms}\equiv\frac{8\pi GD_{\rm eff}}{c^{2}}\rho_{\rms}r_{\rms}$, a dimensionless parameter, denotes a mass surface density. Different halos have different $\kappa_{\rms}$ but the same $\hat{\psi}(x)$ \cite{2004A&A...423..787T}, and low-mass halos have small $\kappa_{\rms}$. In this case, we can define the typical mass of the lens, the dimensionless frequency, and the time delay as
\begin{equation}
M_{Ez}=\frac{r_{\rm s}^2c^2}{4G}\frac{(1+z_\ell)}{D_{\rm eff}},
\end{equation}
\begin{equation}
w=\frac{2\pi f r_{\rm s}^2(1+z_\ell)}{D_{\rm eff} c} = \frac{8\pi G f M_{Ez}}{c^3},
\end{equation}
and 
\begin{equation}
T=cD_{\rm eff}t_{\rm d}/r_s^2.
\end{equation}
Since $r_{\rm s}$ is determined by $M_\ell$ and $z_\ell$, $M_{Ez}$ should depend on $M_\ell$, $z_{\ell}$, and $z_{\rm s}$, from the above description one can obtain 
\begin{equation}
M_{Ez}\propto (r_{\rm vir}/c_{\rm v})^2 \frac{1+z_\ell}{D_{\rm eff}} \propto M_\ell^{0.927} g_1(z_\ell, z_{\rm s}),
\label{eq:MEz}
\end{equation}
where $g_1(z_\ell, z_{\rm s})$ is a function describing the dependence of $M_{Ez}$ on the redshifts of both the GW source and the lens. Similarly, we also have
\begin{equation}
\kappa_{\rm s}\propto D_{\rm eff} \rho_{\rm s} r_{\rm s}\propto M_\ell^{0.463+\frac{d\ln \rho_{\rm s}}{d\ln M_\ell}} g_2(z_\ell, z_{\rm s}),
\label{eq:Kappas}
\end{equation}
where $g_2(z_\ell, z_{\rm s})$ is a function describing the dependence of $\kappa_{\rm s}$ on the redshifts of both the GW source and the lens, and
$$
\frac{d\ln \rho_{\rm s}}{d\ln M_\ell} = \frac{d\ln \rho_{\rm crit}(z_\ell)\delta_{\rm c}(c_{\rm v})}{d\ln M_\ell} \simeq -0.13\frac{d\ln \delta_{\rm c}}{d\ln c_{\rm v}}. 
$$
Here $\rho_{\rm crit}$ is the critical density of universe, $\delta_{\rm c}\equiv\frac{\rho_{\rm s}}{\rho_{\rm crit}}$, \cite{2006glsw.conf.....M} and $\frac{d\ln\delta_{\rm c}}{d\ln c_{\rm v}}\sim 2.67$ for an intermediate lens mass $M_\ell =10^{4.5}M_\odot$ and $z_\ell=2$ and it is gradually changing from $2.62$ to $2.71$ for $c_{\rm v}\in [30, 75]$ (or alternatively $M_\ell \in [10^3,10^6]M_\odot$). Therefore, the power law slope for the dependence of $\kappa_{\rm s}$ on $M_\ell$ is roughly in the range of $0.111-0.123$.

\subsection{The lens density profile and the WDM model}

WDM halos may have a substantial core in the cases of small particle mass (e.g., $m_{\rm p}\lesssim0.5$\,keV) \cite{Lovell:2013ola}, however, many investigations have put constraints on $m_{\rm p}$ to be larger than a few keV \cite{Hsueh:2019ynk}. Here we only consider WDM models with $m_{\rm p}\gtrsim3$\,keV. In these cases, the core size may be small, thus we assume that WDM halos follow the NFW profile as CDM halos do. 

\subsection{The lens density profile and the SIDM model}

SIDM halos may have an isothermal core and follow a piecewise density profile (IC-NFW) as \cite{2016PhRvL.116d1302K}
\begin{equation}
\rho_{\rm SIDM}(x)=\left\{
\begin{array}{ll}{\rho_{\rm iso}(x)}, & {\textrm{for\,} \, 0<x<s}, \\ 
{\rho_{\rm NFW}(x)}, & {\textrm{for\,} \,x\geq s}. 
\end{array}\right.
\label{eq:IC-NFW}
\end{equation}
This profile can be obtained for low-mass halos by solving the Poisson Equation (Eq.~(2) in \citet{2016PhRvL.116d1302K}), assuming the spherical symmetry and neglecting baryonic effects. We find that $s\sim0.1-2$ by assuming self-interaction cross-section $\left<\sigma v\right>/m_{\rm p}\sim1-10^2\,{\rm cm^2\,km\,s^{-1}{\rm g}^{-1}}$ (see  \citet{2016PhRvL.116d1302K}). We consider two cases $s=0.1$ and $s=1$. 

\subsection{Lensing effects}

The differences in density profiles of lenses resulting from different DM models lead to different lensed signals for the same originally unlensed ones. Figures~\ref{fig:F_pJ} and \ref{fig:F_NFW_sCDM} show the amplification factor for some example lenses with the pseudo-Jaffe, NFW, and IC-NFW profiles. As seen from these figures, for the pesudo-Jaffe profile, not only the lensed signals can be significantly different from the original one in both amplitude and phase, but also the signals resulting from lenses with or without core are different from each other. The lensed effects obtained for the pseudo-Jaffe profile also have significant differences compared with those obtained for the NFW or NFW-like profile. These suggest that the lensed GW signals can be used to reconstruct/constrain the lens density profile and thus reveal the DM properties.
Since the difference between the IC-NFW profile and the NFW profile is only at the region within the core size $s$, their amplification factors can be significant different from each other only when $y\lesssim s$.

Here we adopt the methods listed in \cite{Guo2020} to numerically integrate equation~\eqref{eq:AmpFact} and obtain the diffractive lensing effects of GW signals lensed by mini-halos with the NFW/IC-NFW or other profile. We note that \citet{2020ApJ...901...58O} and \citet{2021PhRvD.104f3001C} have considered the diffractive lensing effects by NFW halos under the weak lensing approximation, different from our method. The adoption of the weak lensing approximation may lead to some errors in the lensing effect estimates. \citet{2018PhRvD..98j4029D} and \citet{2022MNRAS.512....1G} calculated the diffractive lensing effects for lenses with the pseudo-Jaffe or SIS profile, which may not represent the real density profile of DM halos.  

\section{SNR analysis of lensed/unlensed GW signals}
\label{sec:SNR}

We define an inner product as
\begin{equation}
\langle g|h\rangle \equiv 4\Re \left( \int_0^{\infty} \frac{\tilde{g}^*(f) \tilde{h}(f)}{S_{\rm n}(f)}df \right),
\end{equation}
where $\Re$ represents the real part, $S_{\rm n}$ the noise power spectral density \cite{2018PhRvD..98j4029D}. Without considering the lensing effect, the SNR of detected signal $s_{\rm gw}$ is  
\begin{equation}
\varrho=|\langle h|s_{\rm gw}\rangle|/\sqrt{\langle s_{\rm gw}|s_{\rm gw}\rangle},
\end{equation}
where $h$ is the best-matched unlensed template. Considering the lensing effect, the difference between lensed and (original) unlensed signals can be described by 
\begin{equation}
\delta\varrho\equiv\sqrt{\langle s_{\rm gw}-h|s_{\rm gw}-h\rangle}
\label{eq:SNRdef}
\end{equation}
\cite{2019PhRvL.122d1103J, 2020MNRAS.495.2002L}, i.e., the `SNR' of difference between two waveforms. The waveforms $s_{\rm gw}$ and $h$ are distinguishable when $\delta\varrho>\delta\varrho_{\rm th}=1$  \cite{Lindblom:2008cm}. Sometimes we adopt $\delta\varrho_{\rm th}=3$ for higher confidence. 

A normalized `SNR' difference can also be defined as $\delta\hat{\varrho} \equiv \sqrt{\langle s_{\rm gw}-h|s_{\rm gw}-h \rangle/ \langle s_{\rm gw}|s_{\rm gw} \rangle}$ \cite{2018PhRvD..98j4029D}. If the unlensed waveform is $\tilde{h}(f)$, we assume that the best fit waveform is $A_{\rm BF}\tilde{h}(f)$, 
\begin{equation}
\delta\hat{\varrho}=\frac{\int_{f_{\min}}^{f_{\max}} \frac{|F(f)-A_{\rm BF}|^2|\tilde{h}(f)|^2} {S_{n}(f)}df }{\int_{f_{\min}}^{f_{\max}} \frac{|\tilde{h}(f)|^2} {S_{n}(f)}df }
\label{eq:dSNR_SNR}
\end{equation}
where $A_{\rm BF}$ is a complex constant that we find the best fit value to make $\delta\varrho$ or $\delta\hat{\varrho}$ smallest. Although phase effect also contribute to $\delta\hat{\varrho}$, we ignore the phase effect in practical calculation and only consider amplitude modulation for convenience. For a lens with the NFW-like profile, the phase of the amplification factor is small ($\lesssim0.01$\,rad), thus the phase effect can be safely ignored. For a lens with the pseudo-Jaffe profile, the phase of the amplification factor may be significant, ignoring the phase effect may lead to a slightly underestimate of the lensing effect and thus the normalized SNR difference $\delta \hat{\varrho}$.

To detect the lensed signal, the lensing effect must be significant enough to satisfy $\delta\hat{\varrho}\geq \delta\varrho_{\rm th}/\varrho_{\rm unl}$, where $\varrho_{\rm unl}$ is the SNR of the original unlensed waveform. 

For convenience, we use a weight function $W(f)=\frac{|\tilde{h}(f)|^2}{S_n(f)}$ to represent the weight of average, thus the Equation~\eqref{eq:dSNR_SNR} becomes
$$
\delta\hat{\varrho}=\frac{\int_{f_{\min}}^{f_{\max}} |F(f)-A_{\rm BF}|^2W(f)df }{\int_{f_{\min}}^{f_{\max}} W(f)df }.
$$
Generally, $W(f)$ depends on the redshifted chirp mass $\mathcal{M}_z=\mathcal{M}_0(1+z_{\rm s})$ ($\mathcal{M}_0$ is the intrinsic chirp mass) and the sensitivity curve of the GW detector. 

The redshifted chirp mass $\mathcal{M}_{z}$ distribution can be obtained by convolving the GW source redshift distribution ($\propto R_{\rm mrg}(z)\frac{dV}{dz}$) and chirp mass $\mathcal{M}_0$ distribution. The resulting $\mathcal{M}_{z}$ distribution has a peak at $\approx22.2M_\odot$, therefore, we adopt the redshifted chirp mass $22.2M_\odot$ to calculate the weight function $W(f)$ for the ground-based GW detectors for simplicity. If adopting a somewhat different redshifted chirp mass, the weight function should have a different shape, but we have checked it would not lead to a significant changes on the magnitude of $\delta\hat{\varrho}$ (the relative difference $<20\%$ from $10M_\odot$ to $100M_\odot$). 
For the middle frequency band, we have $|\tilde{h}(f)|\propto f^{-7/6}$ in the inspiral stage of a sBBH merger, which is independent of the chirp mass. Different sources with different chirp masses have the same weight function $W(f)$ for the same GW detectors. 

In this paper, we consider a number of current and future ground-based high frequency GW observatories and middle frequency space GW detectors. These include LIGO, LIGO A+ \cite{2020LRR....23....3A}, Einstein Telescope (ET) \cite{2012CQGra..29l4006H}, Cosmic Explorer (CE) \cite{2019BAAS...51c.141R}, Gravitational-wave Lunar Observatory for Cosmology (GLOC) \cite{2020arXiv200708550J}), Deci-Hertz Interferometer Gravitational Wave Observatory (DECIGO) \cite{2006CQGra..23S.125K}, and Big Bang Observer (BBO) \cite{2006CQGra..23.4887H}.

\begin{figure}
\centering
\includegraphics[width=0.48\textwidth]{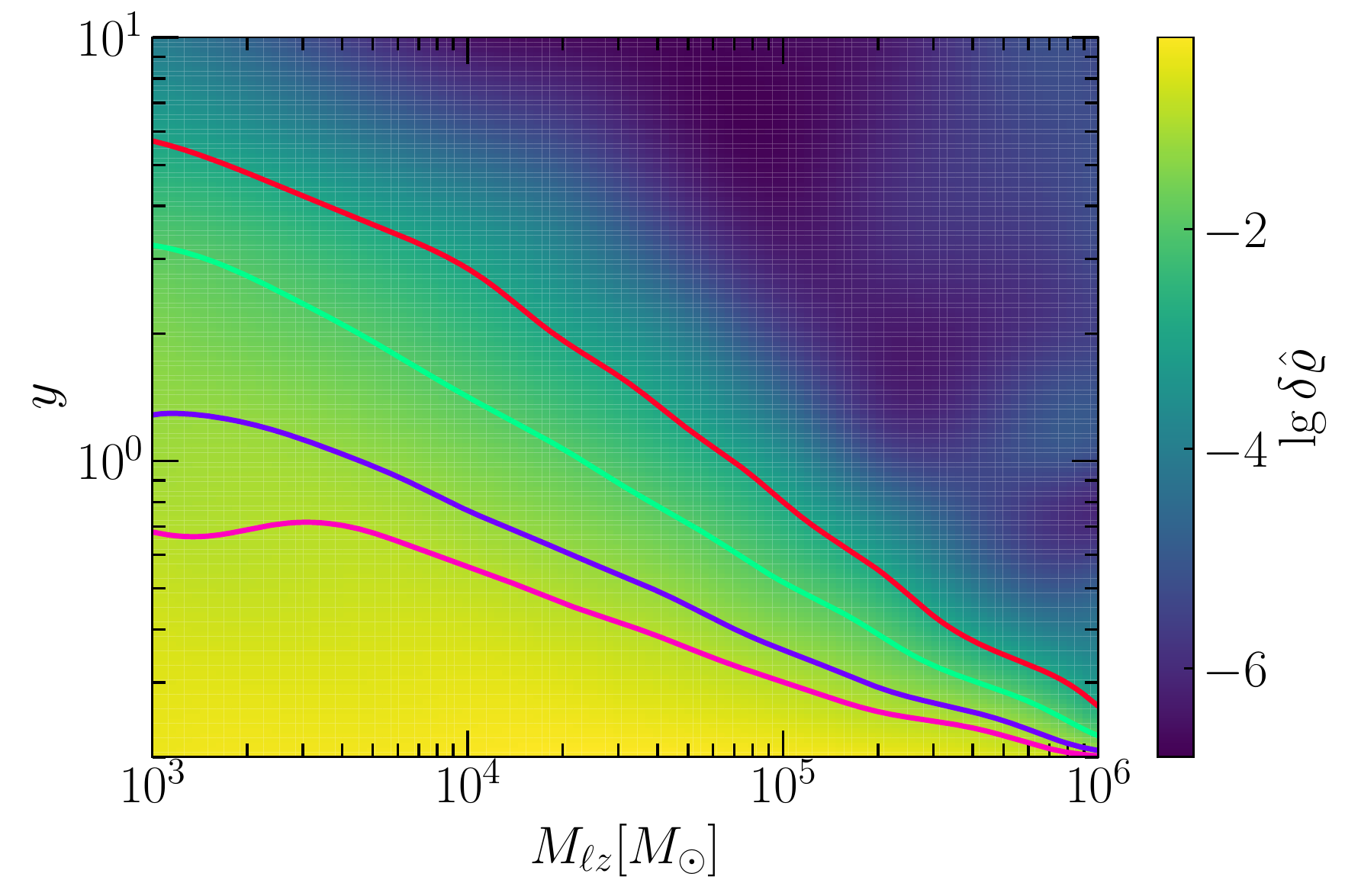} 
\caption{
The distribution of the normalized SNR difference $\delta\hat{\varrho}(M_{\ell z},y)$ on the $y$-$M_{\ell z}$ plane obtained by adopting the GLOC sensitivity curve and assuming the pesudo-Jaffe profile for DM halos. The colorbar at the right side of the figure indicates the value of $\lg\delta \hat{\varrho}$. The solid color contour lines indicate $\delta\hat{\varrho}=0.001$, $0.01$, $0.05$, and $0.1$, respectively (from up to down).
}
\label{fig:dSNR_y_M_pJ}
\end{figure}

\begin{figure}
\centering
\includegraphics[width=0.48\textwidth]{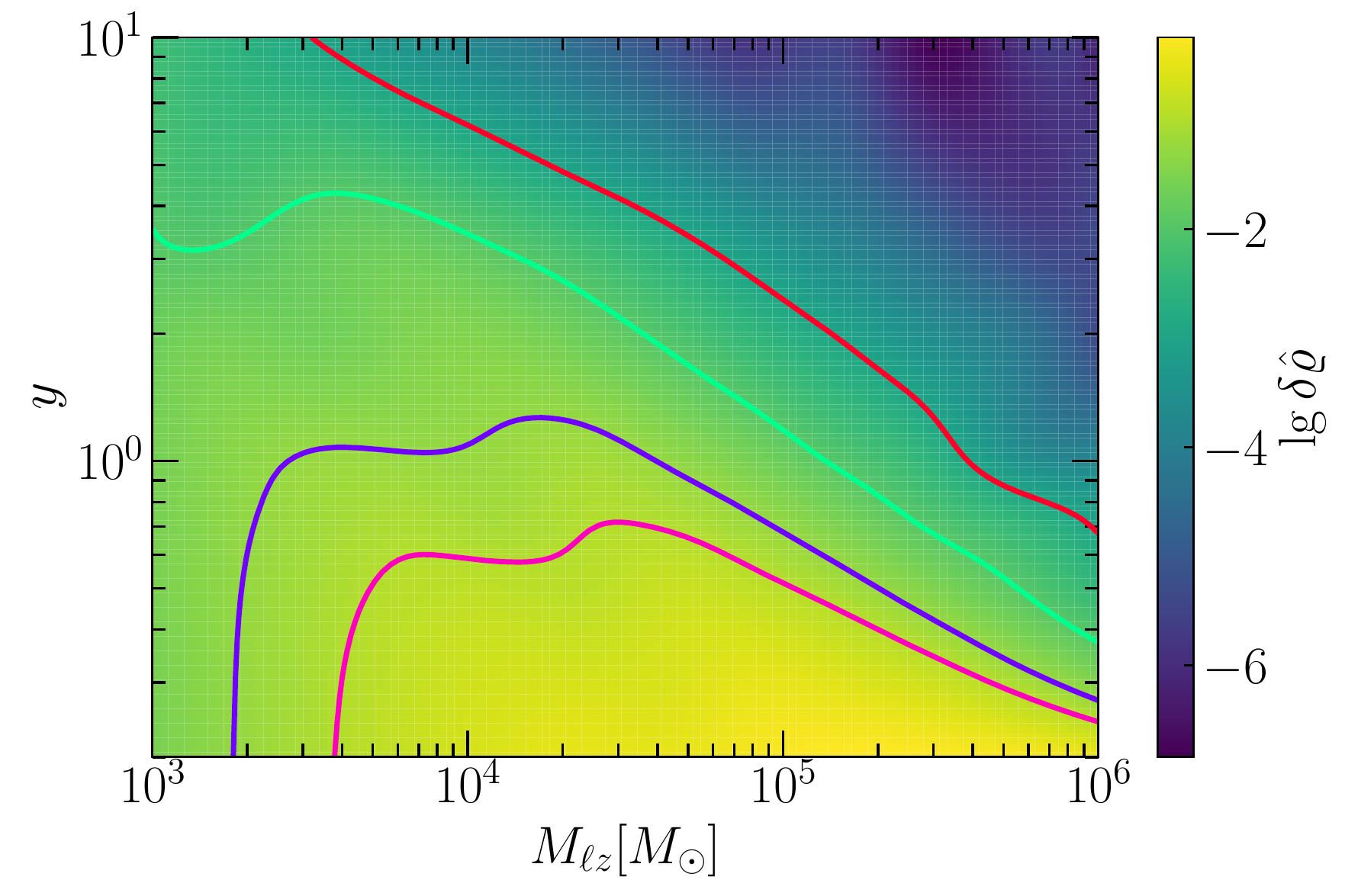} 
\caption{
Legend is similar to that for Fig.~\ref{fig:dSNR_y_M_pJ}, but adopting the BBO sensitivity curve.
}
\label{fig:dSNR_y_M_pJ_BBO}
\end{figure}

Assuming the pseudo-Jaffe profile, $\delta\hat{\varrho}(M_{\ell z},y)$ can be calculated for any given set of ($M_{\ell z}, y$). Figures~\ref{fig:dSNR_y_M_pJ} and \ref{fig:dSNR_y_M_pJ_BBO} show $\delta\hat{\varrho}$ as a function of $y$ and $M_{\ell z}$ ($=M_{\ell}(1+z_\ell))$, obtained by adopting the GLOC and BBO sensitivity curves, respectively. It is obvious that when $y$ is smaller, $\delta\hat{\varrho}$ is greater. For the same $\delta\hat{\varrho}$, $y$ is usually smaller for more massive mini-halos. In the middle frequency band, $y$ will sharply drop when $M_{\ell z}\simeq 10^{3}M_\odot$.
If assuming $\delta \varrho(M_{\ell z}, y_{\rm cr})=\delta\varrho_{\rm th}$ as the threshold for detecting the lensing effect, with which the lensing signal is significant enough to be detected if $y\le y_{\rm cr}$, while it is not if $y>y_{\rm cr}$. Then we can obtain $y_{\rm cr}$ for any given set of ($\varrho_{\rm unl}, M_{\ell z}$) by solving equation $\delta\hat{\varrho}(M_{\ell z},y_{\rm cr})=\frac{\delta\varrho_{\rm th}}{\varrho_{\rm unl}}$, which defines the cross section for the diffractive lensing effect to be significant.

Assuming the NFW or NFW-like profile, $M_{Ez}$ and $\kappa_{\rm s}$ are dependent on as $M_\ell \propto M^{0.926}_\ell$ but dependent on $z_{\rm s}$ and $z_\ell$ in a complicated way (see Eqs.~\ref{eq:MEz} and \ref{eq:Kappas}). For simplicity, we check these dependence by generating a large number of lens systems with realistic distributions of $M_\ell$, $z_\ell$ and $z_{\rm s}$. The probability distribution of $z_{\rm s}$ of sBBH mergers given by the merger rate density distribution $p(z_{\rm s})\propto R_{\rm mrg}(z_{\rm s})\frac{dV}{dz_{\rm s}}$ (see section~\ref{sec:rate} for detailed description). There is no significant difference for the halo mass function (HMF) between $z_{\ell}=0$ and $5$ (see Fig.~\ref{fig:hmf_l} below in Section~\ref{sec:rate}) and the number density of source peaks at $z_{\rm s}\sim2$. Thus we can assume $n_{\ell}(z_\ell) \sim $ constant, for simplicity. The conditional probability distribution for lens is $p(D_\ell|D_{\rm s})dD_\ell\propto n_{\ell}\pi\theta_{\rm E}^2D_\ell^2dD_\ell$, where $\theta_{\rm E}^2\propto\frac{D_{\ell \rm s}}{D_{\rm s}D_{\ell}}$, thus $p(z_\ell|z_{\rm s})\propto n_{\ell}\pi\theta_{\rm E}^2D_\ell^2\frac{dD_\ell}{dz_\ell}=n_{\ell}\pi D_{\rm eff}^2\frac{c}{H(z_\ell)}$. According to $p(z_{\rm s})$ and $p(z_\ell|z_{\rm s})$, we generate $1000$ lens-sources systems with different $M_\ell$ ($\in [10^3,10^{12}]M_\odot$), $z_{\rm s}$ ($\in [0,8])$\footnote{The $z_{\rm s}$ can even extend to 15 in our mocked samples, but when $z>8$, the probability has become so small that can be ignored.} to check the dependence of $M_{Ez}$ and $\kappa_{\rm s}$ on $M_\ell$, $z_{\ell}$ ($\in [0,z_{\rm s}])$, and $z_{\rm s}$. For each set of ($z_{\ell}$, $z_{\rm s}$, $M_{\ell}$), we can calculate the corresponding $M_{Ez}$ and $\kappa_{\rm s}$. For lens halos with the NFW profile and the IC-NFW profile ($s=0.1$ or $s=1$), they roughly have the same $M_{Ez}$ and $\kappa_{\rm s}$ if  ($M_\ell$, $z_\ell$, $z_{\rm s}$) are the same. We find that $M_{Ez}$ and $\kappa_{\rm s}$ mainly depend on $M_\ell$, and the differences in $z_\ell$ and $z_{\rm s}$ of the lens systems introduce some scatters to the main relationship between $M_{Ez}$ (or $\kappa_{\rm s}$) and $M_\ell$. We fit the dependence of either $M_\ell$ or $\kappa_{\rm s}$ on $M_{\ell}$ as a power law with a scatter reflecting the effects of $z_\ell$ and $z_{\rm s}$, and find $M_{Ez}(M_\ell)\approx 28.2M_\odot (M_\ell/M_\odot)^{0.928}$ with a scatter of $\sigma_{\lg M_{Ez}/M_\odot}=0.221$, and $\kappa_{\rm s} \approx 0.00510 (M_\ell/M_\odot)^{0.121}$ with a scatter of $\sigma_{\lg\kappa_{\rm s}}=0.136$. Apparently, the index $0.928$ is consistent with the estimate in equation~\eqref{eq:MEz}, and the redshift distributions of $z_\ell$ and $z_{\rm s}$ only lead to a small scatter (the factors $g_1(z_{\ell},z_{\rm s})$ and $g_2(z_{\ell},z_{\rm s})$ in Eqs.~\ref{eq:MEz} and \ref{eq:Kappas}) to the relationship between $M_{Ez}$ and $M_\ell$. For simplicity, we thus ignore the dependence on $z_\ell$ and $z_{\rm s}$ for when estimating the cross section for those ``detectable'' lensed events below.

\begin{figure}
\centering
\includegraphics[width=0.48\textwidth]{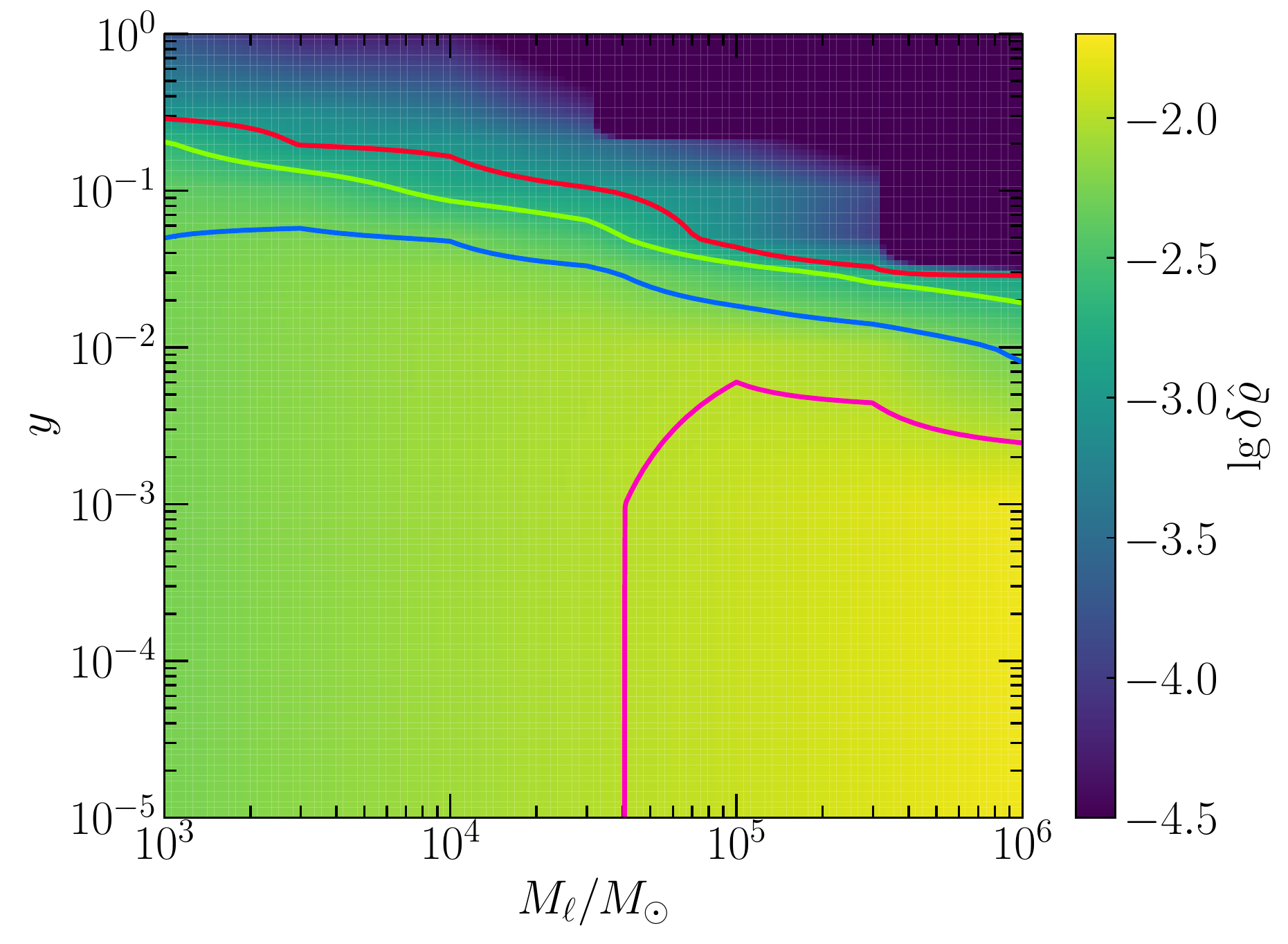}  
\caption{
Legend is similar to that for Fig.~\ref{fig:dSNR_y_M_pJ}, but adopting the NFW density profile for mini-halos and the GLOC sensitivity curve. The solid color contour lines indicate $\delta\hat{\varrho}=0.001$, $0.002$, $0.005$ and 0.01, respectively (from top to bottom).
}
\label{fig:dSNR_y_NFW}
\end{figure}

\begin{figure}
\centering
\includegraphics[width=0.48\textwidth]{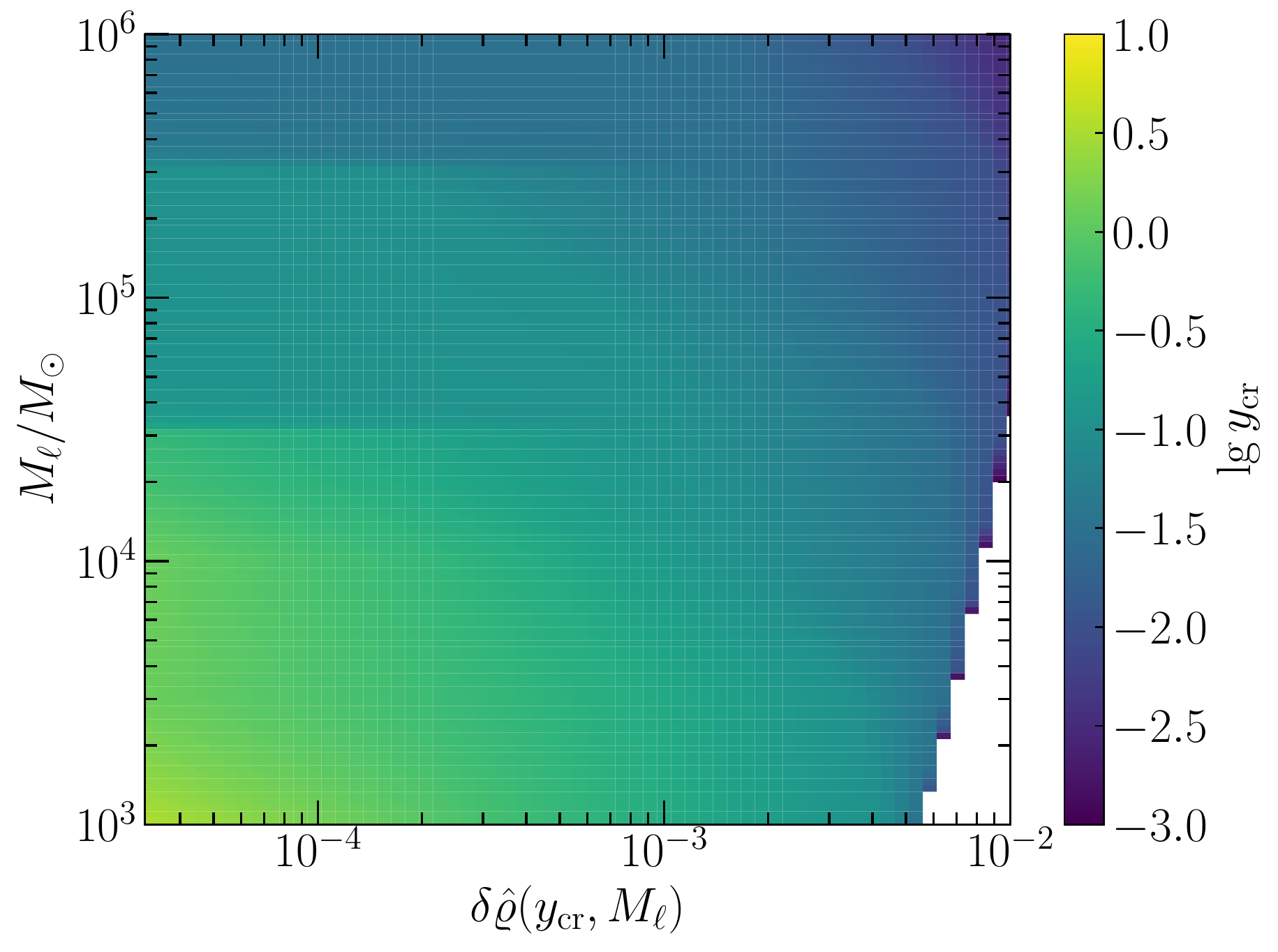}  
\caption{
The ``detectable'' threshold $y_{\rm cr}$ of the diffractive lensing effect by mini-halos with $M_\ell$ ($\in [10^3,10^6]M_\odot$) and the SNR difference of $\delta\hat{\varrho}$ obtained by adopting the NFW profile for mini halos and the GLOC sensitivity curve. The blank region represents $y_{\rm cr} \rightarrow0$.
}
\label{fig:dSNR_yM_NFW}
\end{figure}

\begin{figure}
\centering
\includegraphics[width=0.48\textwidth]{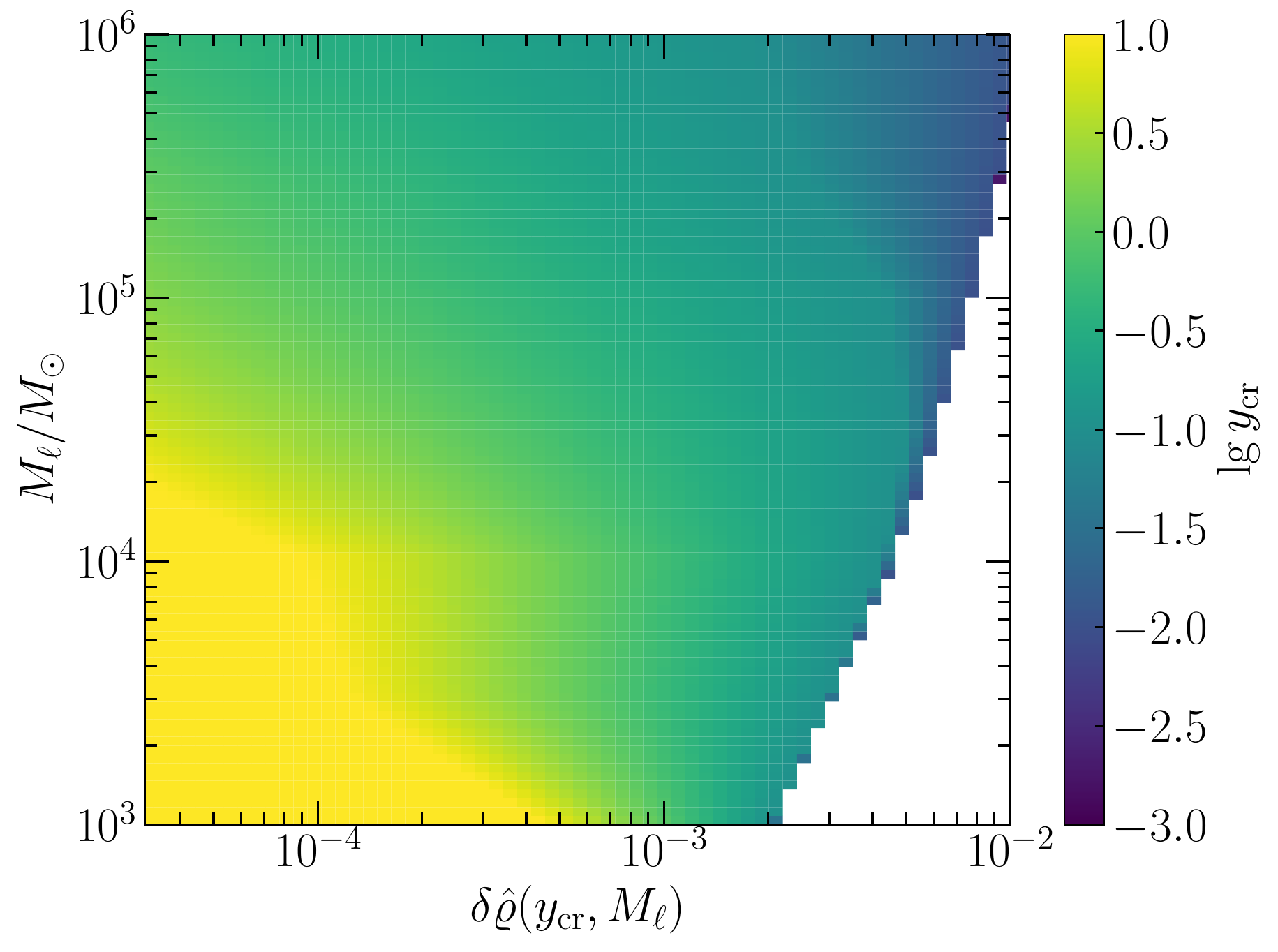}  
\caption{
Legend is similar to that for Fig.~\ref{fig:dSNR_yM_NFW} except that the BBO sensitivity curve is adopted.
}
\label{fig:dSNR_yM_NFW_Mid}
\end{figure}

For mini-halo lenses with the NFW (or NFW-like) profiles, $\delta\hat{\varrho}$ can be estimated, as a function of halo mass $M_\ell$ ($\in [10^3,10^{6}]M_\odot$) and $y$ ($\in[10^{-5},1]$). We adopt the above fitting relationships between $M_{Ez}$ (or $\kappa_{\rm s}$) and $M_\ell$ without considering the scatters due to the distributions of $z_{\ell}$ and $z_{\rm s}$, for simplicity.
If we set the SNR difference $\delta\varrho_{\rm th}= 1$ as the threshold for ``detectable'' lensed events, then we can obtain the threshold for $y$ parameter, i.e., $y_{\rm cr}$, by solving equation $\delta\hat{\varrho}(y_{\rm cr},M_\ell)=\delta\varrho_{\rm th}/\varrho_{\rm unl}$. Figure~\ref{fig:dSNR_y_NFW} shows $\delta\hat{\varrho}$ as a function of $y$ and $M_{\ell}$ for the NFW profile by adopting GLOC sensitivity curve as an example. Figures~\ref{fig:dSNR_yM_NFW} and \ref{fig:dSNR_yM_NFW_Mid} show the distributions of $y_{\rm cr}(\delta\hat{\varrho},M_\ell)$ on the $M_\ell$-$\delta\hat{\varrho}$ plane for the NFW profile by adopting the GLOC and BBO sensitivity curves, as examples for the high-frequency ground-based GW observatories and the mid-frequency space GW detectors, respectively. In both figures, the smaller lens mass and the smaller $\delta \hat{\varrho}(y_{\rm cr},M_\ell)$, the relatively larger $y_{\rm cr}$. By comparison of these two figures, it can be clearly seen that $y_{\rm cr}$ obtained for the BBO sensitivity curve is substantially larger than that for the GLOC sensitivity curve because the diffraction effect is more significant in the middle-frequency band than that in the high-frequency band for those mini-halos. Generally, the $y_{\rm cr}$ is greater for smaller $\delta\hat{\varrho}$.

\begin{figure}
\centering
\includegraphics[width=0.48\textwidth]{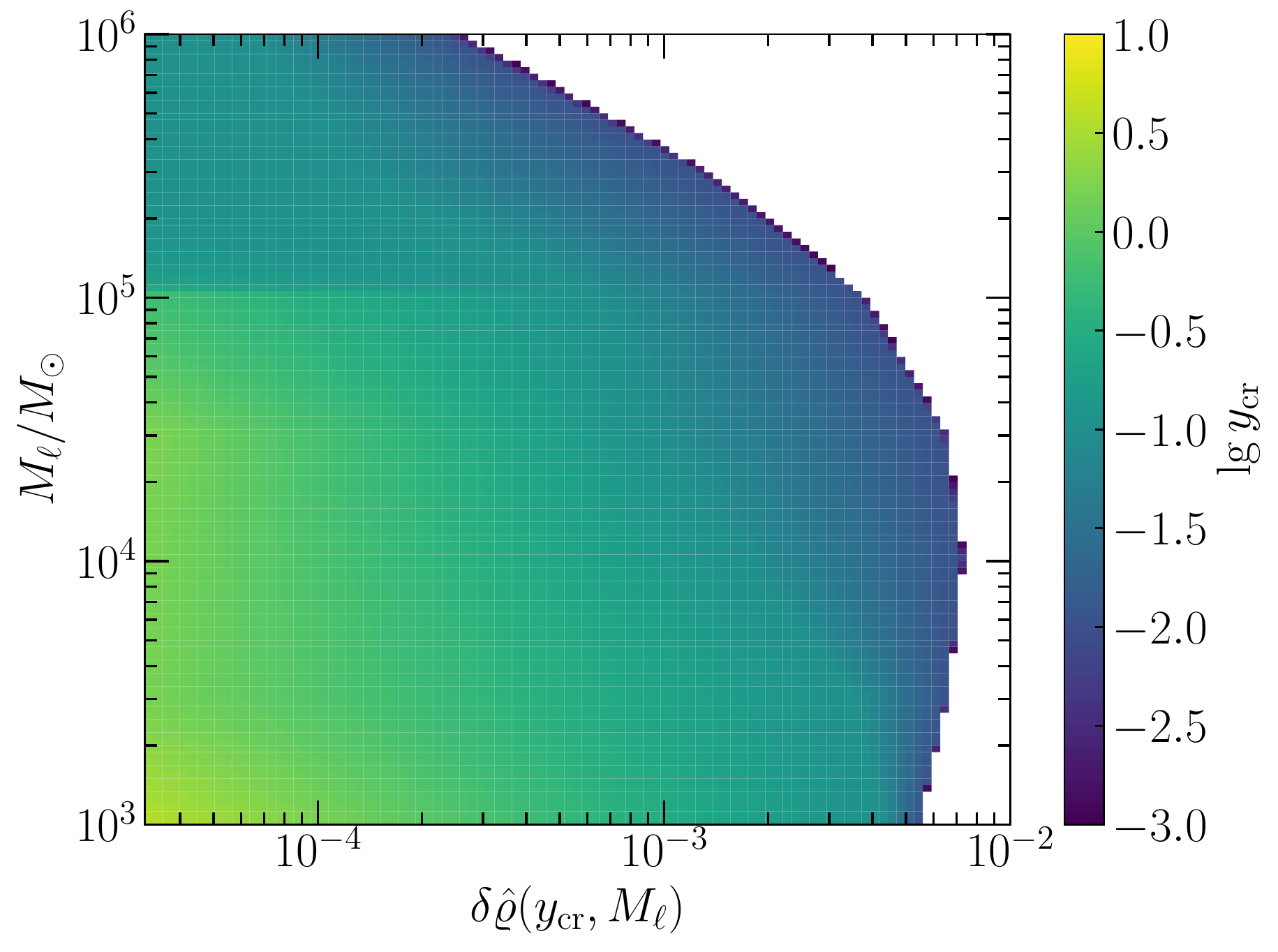}  
\caption{
Legend is similar to that for Fig.~\ref{fig:dSNR_yM_NFW} except that the SIDM density profile is adopted.
}
\label{fig:dSNR_yM_sCDM}
\end{figure}

\begin{figure}
\centering
\includegraphics[width=0.48\textwidth]{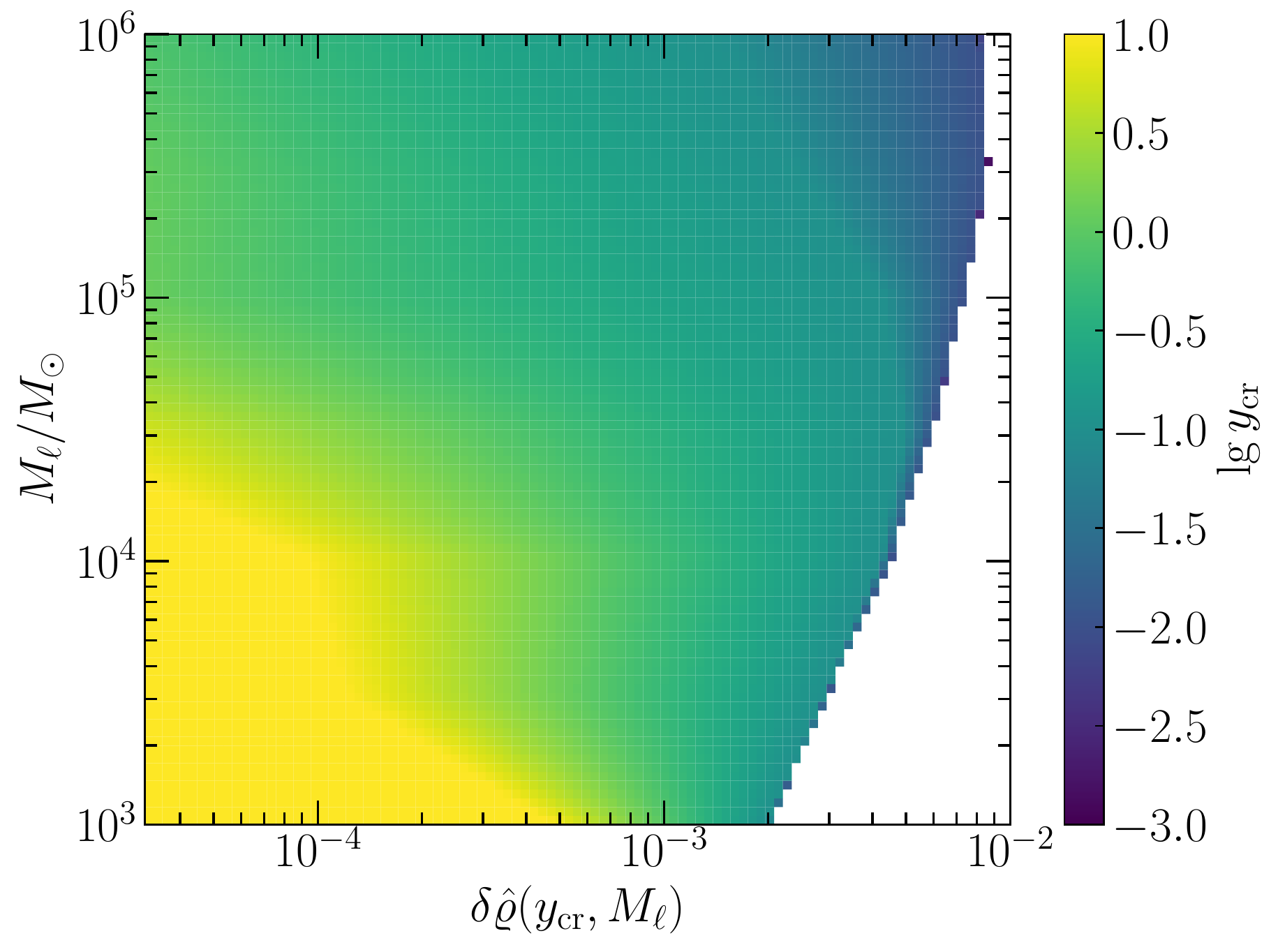}  
\caption{
Legend is similar to that for Fig.~\ref{fig:dSNR_yM_NFW_Mid} except that the SIDM density profile is adopted.
}
\label{fig:dSNR_yM_sCDM_Mid}
\end{figure}

For the IC-NFW profile with $s=0.1$ and $s=1$, we also calculate $\delta\hat{\varrho}$ for different halo masses $M_\ell$ ($\in [10^3,10^6]M_\odot$) and $y$ ($\in [10^{-5},1]$).  
Figures~\ref{fig:dSNR_yM_sCDM} and \ref{fig:dSNR_yM_sCDM_Mid} show the distributions of $y_{\rm cr}(\delta\hat{\varrho},M_\ell)$ for mini-halos with the IC-NFW density profile with $s=0.1$ on the $\delta\hat{\varrho}$-$M_\ell$ plane by adopting the GLOC and BBO sensitivity curves, respectively. 
For the ground-based GW detectors in the high frequency band, $y_{\rm cr}$ for mini-halos with the IC-SIDM profile (Fig.~\ref{fig:dSNR_yM_sCDM}) is significantly smaller than that for those with the NFW profile (Fig.~\ref{fig:dSNR_yM_NFW}), especially at large $M_\ell\sim10^6M_\odot$. The reason for this is that the amplification factors of those mini-halo lenses with the IC-NFW profile and the NFW profile have significant difference in the high-frequency band. The former ones become flat when $f$ (or $w$) is sufficiently large for small $y\lesssim s$ (see Fig.~\ref{fig:F_NFW_sCDM}), while the latter ones still increase with increasing $f$ (or $w$).
This is also the reason why the lensing rate of SIDM is smaller than CDM with NFW profile for ground-based detectors that will be introduced in later sections. However, for the space GW detectors in the middle-frequency band, $y_{\rm cr}$ for mini-halos with the IC-NFW profile (Fig.~\ref{fig:dSNR_yM_sCDM_Mid}) is  similar to that for those with the NFW profile with $s=0.1$ (Fig.~\ref{fig:dSNR_yM_NFW_Mid}). The reason is that the amplification factors for mini-halos with the NFW profile and those with the IC-NFW profile ($s=0.1$) are almost the same in the middle-frequency band even when $y$ is small. Therefore, their lensing rates are also similar to each other as shown in section~\ref{sec:rate}. If the IC-NFW profile has a larger core, e.g., $s=1$, the resulting $y_{\rm cr}$ will be significantly smaller than that from the NFW profile at both high-frequency and middle frequency bands.  

The cross-section for mini-halos in the wave optics regime  is substantially larger than that in the geometrical limit (zero or very small depending on the density profile and lens masses; see \cite{2020MNRAS.497.4956J}), thus the diffractive lensing rate of GW sources by mini-halos is much larger than the corresponding strong lensing rate in the geometrical optics regime. Here we ignore the strong lensing effect as its cross section for mini-halos is negligible comparing with that of the diffractive lensing effect. We adopt the critical value $y_{\rm cr}$ to infer the cross section $\sigma=\pi (\xi_0(M_\ell) /D_\ell)^2 y_{\rm cr}^2$ and calculate diffractive lensing event rate.

Given the lensing cross-sections, the optical depth for a GW source at $z_{\rms}$ lensed by halos is
\begin{equation}
\tau\left(\varrho,z_{\rms}\right)=\frac{1}{4\pi}\int_0^{z_{\rms}} dz_\ell\int dM_{\ell} \pi (\xi_0(M_\ell)/D_\ell)^2 y_{\rm cr}^2  \frac{dn}{dM_\ell } \frac{{\rm d}V^c}{{\rm d}z_\ell},
\end{equation}
where $dn/dM_\ell$ is the HMF and can be calculated using the python program \texttt{hmf} \cite{2013A&C.....3...23M}, $V^{\rm c}$ is the cosmic comoving volume, the integration ranges for $z_\ell$ and $M_\ell$ are $(0,z_{\rms})$ and $(10^3,10^6)M_\odot$, respectively. For lower mass halos, the lensing effect is substantially weaker, while for higher mass halos, they may not act in the diffraction regime and their density profiles may be significantly affected by baryonic processes. 

\begin{table*}
\caption{Expected diffraction lensing rate $\dot{N}_\ell$(yr$^{-1}$) of stellar sBBH mergers/inspiral for LIGO, LIGO A+, ET, CE, GLOC, DECIGO, and BBO, estimated by assuming different DM models (CDM; WDM with particle mass of $30$, $10$, and $3$keV, respectively; and SIDM), and different density profiles (pseudo-Jaffe, NFW, and IC-NFW) for halos. The second column indicates the expected total sBBH detection rates by different GW detectors with a SNR threshold of $\varrho_{\rm th}=8$. The last row (for BBO) and the last third row (for DECIGO) show the results obtained by assuming a conservative SNR difference threshold of $\delta\varrho_{\rm th}=3$, while other rows show the results obtained by assuming $\delta\varrho_{\rm th}=1$.
}
\begin{center}
\begin{tabular}{c|c|c|ccc|c|ccc|cc} \hline\hline
\multirow{4}{*}{Detector} & \multirow{4}{*}{$\dot{N}_{\rms}[\rm yr^{-1}]$} & \multicolumn{9}{c}{$\dot{N}_\ell$[yr$^{-1}$]}\\ \cline{3-12}
& & \multicolumn{4}{c}{pseudo-Jaffe} \vline & \multicolumn{4}{c}{NFW } \vline &\multicolumn{2}{c}{IC-NFW } \\     \cline{3-12}
&  & \multirow{2}{*}{CDM} & \multicolumn{3}{c}{WDM} \vline & \multirow{2}{*}{CDM} & \multicolumn{3}{c}{WDM} \vline  & \multicolumn{2}{c}{SIDM} \\   \cline{4-6} \cline{8-12}
& & & 30keV & 10keV & 3keV & &30keV & 10keV & 3keV & $s=0.1$ & $s=1$  \\ \hline
%
%
LIGO 
& 971 
& 0.344 & 0.0688 & 0.0133 & 0.00143 &
$1.85$$\times$$10^{-7}$  & $5.09$$\times$$10^{-8}$ & $9.53$$\times$$10^{-9}$ & $1.01$$\times$$10^{-9}$ 
& $8.52\times$$10^{-8}$ & $6.18\times$$10^{-10}$ \\ 
LIGO A+ 
& 2502 
& 1.41 & 0.303 & 0.0599 & 0.00640 &
$6.44$$\times$$10^{-7}$ & $1.80$$\times$$10^{-7}$ & $3.38$$\times$$10^{-8}$ & $3.58$$\times$$10^{-9}$ 
& $2.88$$\times$$10^{-7}$  & $2.45$$\times$$10^{-9}$ \\ 
ET 
& $3.26$$\times$$10^4$ & 207 & 31.7 & 5.49 & 0.563 & 
0.00421 & 0.00138 & $2.64\times$$10^{-4}$ & $2.78$$\times$$10^{-5}$ 
& 0.00125  & $8.93\times$$10^{-6}$ \\
CE 
& $3.82$$\times$$10^4$ & 274 & 37.8 & 6.47 & 0.661 & 
0.288 & 0.0899 & 0.0169 & 0.00172 
& 0.0980  & $4.69$$\times$$10^{-4}$ \\
GLOC 
& $3.90$$\times$$ 10^4$ 
& 893 & 115 & 18.2 & 1.84 &
1.02 & 0.243 & 0.0437 & 0.00436
& 0.585 & 0.00223 \\
\hline
\multirow{2}{*}{DECIGO} 
& \multirow{2}{*}{$3.92$$\times$$10^{4}$}
& 3640 & 516 & 77.6 & 7.76 &
9.67 & 3.25 & 0.601 & 0.0604
& 9.61  & 2.37 \\
& & 1112 & 193 & 30.7 & 3.09 &
0.0778  & 0.0306 & 0.00634 & $6.15$$\times$$10^{-4}$
&  0.0745 & 0.0132
\\ \hline
\multirow{2}{*}{BBO} 
& \multirow{2}{*}{$3.97$$\times$$10^4$} 
& 8022 & 996 & 146 & 14.5 &
189 & 44.1 & 7.81 & 0.803
& 192 & 88.2 \\
& & 3281 & 448 & 66.9 & 6.68 &
5.66  & 1.96 & 0.367 & 0.0376
& 5.48  & 0.906
\\

\hline \hline 
\end{tabular}
\end{center}
\label{tab:eventrate}
\end{table*}

\section{Expected lensing rate}
\label{sec:rate}

After defining the threshold of cross section $y_{\rm cr}$ from SNR analysis, we can calculate the expected lensing rate for current and future GW detectors at the high frequency band (Sec.~\ref{sec:high}) and the middle frequency band (Sec.~\ref{sec:mid}), respectively. Finally, our results are presented in Sec.~\ref{sec:res}.

\subsection{High frequency band}
\label{sec:high}

The differential event rate of GW sources (e.g., sBBH mergers) is given by
\begin{equation}
\frac{d^{3}\dot{N}}{dzd\mathcal{M}d\varrho}=P_{\varrho}\left(\varrho|z,\mathcal{M}_{0}\right)\frac{R_{\mathrm{mrg}}\left(\mathcal{M}_{0};z\right)}{(1+z)} \frac{dV^{\rm c}}{dz},
\end{equation}
where $R_{\mathrm{mrg}}\left(\mathcal{M}_{0};z\right)=p(\mathcal{M}_{0}|z)R_{\mathrm{mrg}}(z)$ is the merger rate density (MRD) for events with chirp mass $\mathcal{M}_0$ at $z$ \cite{2018MNRAS.476.2220L}, and $P_{\varrho}(\varrho|z,\mathcal{M}_0)$ is the conditional probability distribution of SNR at $z$ and  $\mathcal{M}_0$ \cite{2018MNRAS.476.2220L,1996PhRvD..53.2878F}. 
We estimate the sBBH MRD by using the ``R3:1'' model in \cite{2021MNRAS.500.1421Z}, with which $75\%$ sBBHs are formed from evolution of field binary stars and the rest originated from dynamical interactions, but re-scale it to the latest constraint on the local MRD \cite{LIGO_O3a}. This model is adopted as its resulting MRD evolution and chirp mass distribution are more or less consistent with current LIGO/Virgo observations. There are still some uncertainties in the constraints on the MRD and its evolution \cite{LIGO_O3a}, which directly introduce errors into the lensing rate estimates, but by a factor $<2$.

The detection rate of lensed GW events is
\begin{equation}
\dot{N}_{\ell}\left(>\varrho_{\rm th}\right) \simeq \int_{\varrho_{\rm th}}^{\infty} d\varrho \iint dz_{\mathrm{s}} d\mathcal{M}_0  p_\ell\left(\varrho, z_{\mathrm{s}}\right) \frac{d^3\dot{N}}{dz_{\rms} d\mathcal{M}_0 d\varrho},
\label{eq:lensrate}
\end{equation}
where the lensing probability $p_\ell(\varrho,z_{\rms})=1-e^{-\tau(\varrho,z_{\rms})}$. The total detection rate of GW events is
\begin{equation}
\dot{N}_{\rms}\left(>\varrho_{\rm th}\right)\simeq\int_{\varrho_{\rm th}}^{\infty}d\varrho\iint dz_{\mathrm{s}}d\mathcal{M}_0   \frac{d^{3}\dot{N}}{dz_{\rms}d\mathcal{M}_{0}d\varrho}.
\end{equation}
Here we ignore the magnification bias as the amplification amplitude is close to $1$ for most lensed events in the wave optics regime.

\subsection{Middle frequency band}
\label{sec:mid}

For the middle-frequency band GW detectors, the target GW sources are no longer the sBBH mergers, but the inspiraling sBBHs. They are continuous GW sources and will last for a long time. The differential event number is \citep{2021MNRAS.500.1421Z,2021RAA....21..285C}  
%
\begin{eqnarray}
%
%
\frac{d^3N}{dzd\mathcal{M}_0d\varrho}& = & P_{\varrho}\left(\varrho|z,\mathcal{M}_{0}\right)\frac{R_{\rm mrg}(\mathcal{M}_0,z)}{(1+z)}\frac{dV^c}{dz} \nonumber \\
& & \cdot\int_{\log_{10} f_{\min}}^{\log_{10} f_{\max}}\frac{dt}{d\log_{10} f}d\log_{10} f,
%
%
\label{eq:dNmid}
%
\end{eqnarray}
where 
$$
\frac{d t}{d \log_{10} f}
=\frac{5 \ln 10}{96} \pi^{-8 / 3}\left[\frac{G \mathcal{M}_0(1+z)}{c^{3}}\right]^{-5 / 3} f^{-8 / 3}.
$$
For the middle frequency GW detectors, the targeted frequency range is normally from $0.1$ to $10$\,Hz. In Equation~\eqref{eq:dNmid}, we simply set $f_{\max}=10$\,Hz, and the set a slightly different value for $f_{\rm max}$ does not affect the integral value much as the residence timescale for sBBHs $\propto f^{-8/3}$. For a given observation period $\Delta t$, those events that can sweep over the targeted frequency range of the middle frequency GW detectors can have the frequency at the start time of the observation as 
$$
f_{\min}^{-\frac{8}{3}}-f_{\max}^{-\frac{8}{3}}=8 \cdot \frac{32}{5} \pi^{\frac{8}{3}}\left(\frac{G \mathcal{M}_z}{c^{3}}\right)^{\frac{5}{3}}\Delta t,
$$
and thus
$$
f_{\min}\approx 0.0335\rm Hz\left(\frac{\mathcal{M}_z}{22.2M_\odot}\right)^{\frac{5}{8}}\left(\frac{\Delta t}{1\rm yr}\right)^{-\frac{3}{8}}.
$$
If $\Delta t>1$\,yr, we have $f_{\rm min}<0.1$\,Hz, which means that we can also roughly define the detection rate as $\frac{1}{\Delta t} d^3N/dzd\mathcal{M}_0d\varrho$, and the SNR can be estimated as
$$
\varrho=8\Theta \frac{r_{\rm det}}{d_{\rm L}}\left(\frac{\mathcal{M}_z}{1.2 M_{\odot}}\right)^{5 / 6}.
$$
Here $d_{\rm L}$ is the luminosity distance, $\Theta$ is the orientation function \citep{1996PhRvD..53.2878F}, and we denote
$$
r_{\rm det}^{2} \equiv \frac{5}{192 \pi}\left(\frac{3G}{20}\right)^{5 / 3} \frac{M_{\odot}^{2}}{c^3}\int_{f_{\min,\rm det}}^{f_{\max,\rm det }} \frac{\left(\pi M_{\odot}\right)^{2}}{\left(\pi f M_{\odot}\right)^{7 / 3} S_{\rm n}(f)} d f,
$$
with $f_{\max,\rm det}=10$\,Hz and $f_{\min,\rm det}=0.01$\,Hz. We adopt
the sensitivity curves $S_{\rm n}(f)$ for DECIGO and BBO as those  given in \citep{2011PhRvD..83d4011Y}.

\subsection{Results}
\label{sec:res}

\begin{figure*}
\centering
\includegraphics[width=0.9\textwidth]{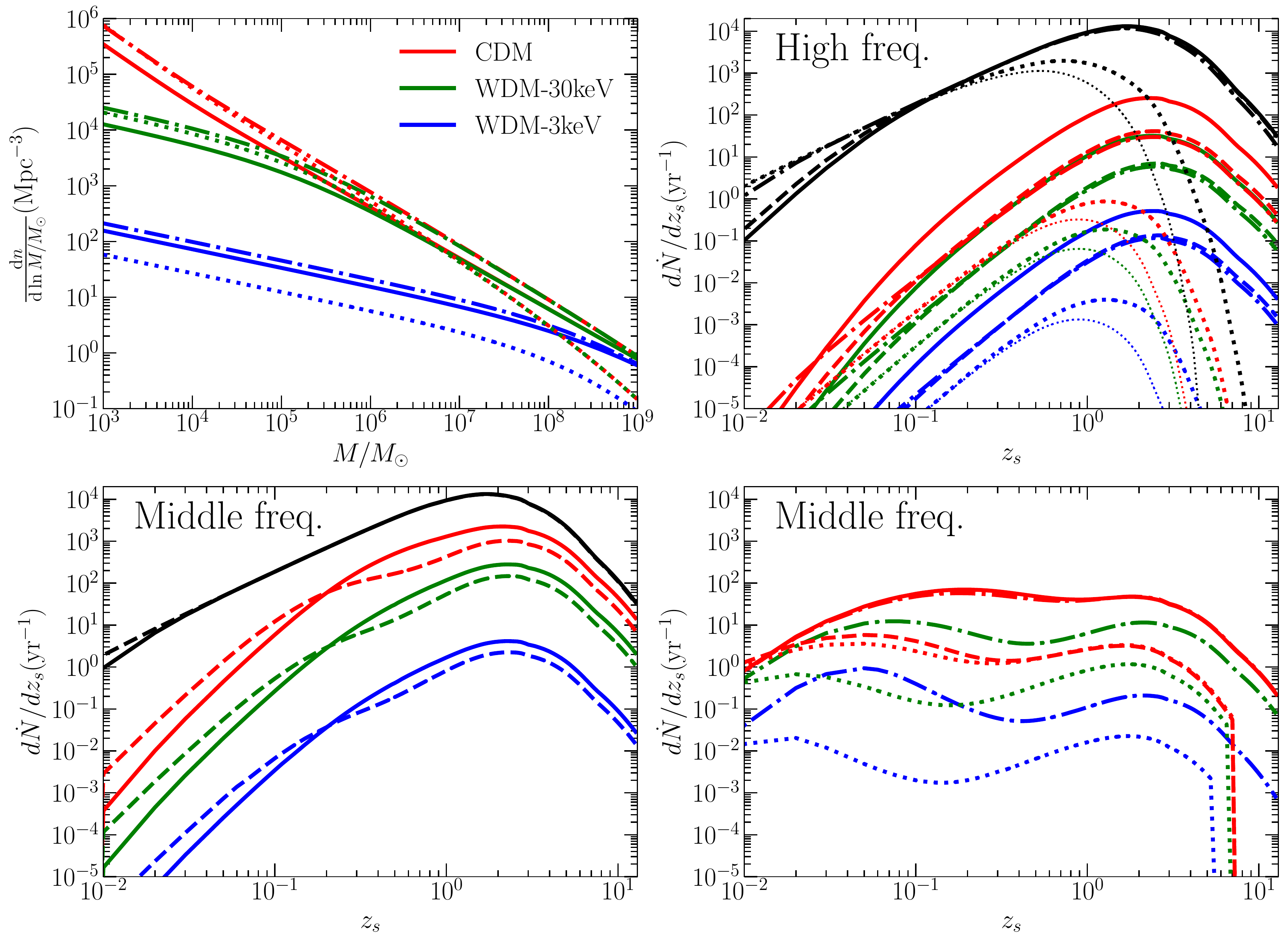} 
%
%
\caption{The halo mass function (top left panel) and the differential diffractive lensing rates of sBBH mergers (other panels) estimated for different DM models. Red, green, and blue lines represent those obtained from the models assuming CDM, WDM with mass of $30$\,keV (WDM-30keV), and $3$\,keV (WDM-3keV), respectively. 
Top left panel: the solid, dotted-dashed, and dotted lines show the halo mass functions at $z=0$, $5$, and $10$, respectively. 
Top right panel: the black lines show the total GW event rates; the thin dotted, thick dotted, thick dashed, thick dotted-dashed, and thick solid lines represent those obtained by using the LIGO, LIGO A+, ET, CE, and GLOC sensitivity curves, respectively. The red, green, and blue lines represent the differential lensing rates obtained from the CDM, WDM-30keV, and WDM-3keV models, respectively. For illustration, the pseudo-Jaffe profile is adopted for all DM halos (with $(s, a)=(0.1, 2)$) for the results showing in this panel. 
Bottom left panel: the black lines shows the total GW event rate in the middle frequency band; the solid and dashed lines represent the differential lensing rates obtained by using the BBO and DECIGO sensitivity curves, respectively, and the DM halos are also assumed here to follow the pseudo-Jaffe profile.   
Bottom right panel: the red, green, and blue lines represent the estimated differential lensing rates for the CDM/SIDM, WDM-30keV, and WDM-3keV model, respectively, where the red dotted-dashed line and the red dotted line represent the results for the CDM model, and the red solid line and the red dashed line represent the results for the SIDM model. The solid line and the dotted-dashed line represent the rates obtained by using the BBO sensitivity curve; the dashed line and the dotted line both represent those obtained by using the DECIGO sensitivity curve. 
For this panel, the DM halos are assumed to follow the NFW profile in the CDM and WDM models, but an IC-NFW profile with $s=0.1$ (see Eq.~\ref{eq:IC-NFW}) in the SIDM model.
}
\label{fig:hmf_l}
\end{figure*}

We estimate the number of ``detectable'' lensed GW events for different GW detectors including the ground-based GW observatories (LIGO, LIGO A+, ET, CE, and GLOC) and the middle frequency GW detectors (DECIGO and BBO), by assuming different DM models and halo density profiles. Our results are listed in Table~\ref{tab:eventrate} and shown in Figure~\ref{fig:hmf_l}. Apparently, the predicted ``detectable'' lensing event rates for a given GW detector can differ by orders of magnitude if assuming different DM models, because of the differences in both the resulting halo abundance and density profiles. As seen from the top-left panel of Figure~\ref{fig:hmf_l}, the HMF at the low-mass end ($\sim10^3-10^6M_\odot$) resulting from the WDM model (with $m_{\rm p}\lesssim30$\,keV) is more than one to two orders of magnitude smaller than that from the CDM model. Other panels of Figure~\ref{fig:hmf_l} show the differential ``detectable'' lensing rates by different GW detectors. For the third generation GW detectors ET, CE, and GLOC, and the middle frequency GW detectors DECIGO and BBO, it seems that the expected rate of all the ``detectable'' GW events does not increase significantly with increasing sensitivity, mainly caused by that most GW events at redshift $z\lesssim 5$ will be ``detected'' by these GW detectors as they are all sufficiently sensitive. Nevertheless, the SNRs of all those ``detectable'' GW events will be significantly enhanced with increasing GW detection sensitivity. Therefore, those GW detectors with high sensitivity can detect more lensed GW events because the lensed signal can be easier identified via the SNR difference threshold.

If DM is cold, the detection rate of GW events lensed by CDM mini-halos with NFW profiles is estimated to be  $\sim 1.85\times10^{-7}$, $6.44\times10^{-7}$, $0.00421$, $0.288$, $1.02$, $9.67$, and $189$\,yr$^{-1}$ for LIGO, LIGO A+, ET, CE, GLOC, DECIGO, and BBO, respectively, while it is $\sim 0.344$, $1.41$, $207$, $274$, $893$, $3640$, and $8022$\,yr$^{-1}$, respectively, if assuming the pseudo-Jaffe profile. The reason is that the pseudo-Jaffe halos are more concentrated than those NFW ones, and thus have relatively larger cross-sections for diffractive lensing. According to these estimates, it is expected that tens to hundreds of such lensed events may be detected 
in the era of ET/CE/GLOC, and even more than thousands of lensed events can be detected 
for DECIGO and BBO. 

If DM is warm with $m_{\rm p}=30$\,keV, then this detection rate is $\sim 5.09\times10^{-8}$, $1.80\times10^{-7}$, $0.00138$, $0.0899$, $0.243$, $3.25$, and $44.1$\,yr$^{-1}$ (or $0.0688$, $0.303$, $31.7$, $37.8$, $115$, $516$, and $996$\,yr$^{-1}$) for LIGO, LIGO A+, ET, CE, GLOC, DECIGO, and BBO, respectively, by adopting the NFW (or pseudo-Jaffe) profile. These expected rates are substantially smaller than those from the CDM model mainly because the abundance of mini-halos resulting from the WDM model is substantially smaller than that from the CDM model. Assuming the WDM model with $m_{\rm p} \gtrsim 10$keV, DECIGO (BBO) is expected to detect about more than one (several) lensed GW events per year. However one would not expect to detect any diffractively lensed GW event by GLOC and DECIGO within a reasonable observation period, if $m_{\rm p}$ is much smaller ($\ll 10$\,keV).

As for the SIDM model, the expected lensing rate could be different for different core size $s$. If the core size $s=0.1$, the expected lensing rate detected by DECIGO/BBO is similar to that expected from the CDM model because the lensing effects of the IC-NFW halos is almost the same as that of NFW lens. But if we have detected a lensed event with $y \lesssim 0.1$, it is still possible to distinguish different lens profiles by the lensed waveform. For the SIDM model, if the lenses follow the IC-NFW profile with large $s$, e.g., $s=1$, the lensing rate could be significantly less than that of the NFW lens or the IC-NFW lens with $s=0.1$. The significant difference between the lensed GW waveforms by the IC-NFW halos and that of the NFW halos can also be used to distinguish the SIDM model with $s=1$ from the CDM model (Fig.~\ref{fig:F_NFW_sCDM}).

Note that we set the threshold for the lensed GW signatures to be identifiable as $\delta\varrho \geq\delta\varrho_{\rm th}=1$ above. However, one may also set a more conservative threshold, for which the estimate lensing rates should decrease substantially and the confidence for the defined ``detectable'' event would be much higher. In Table~\ref{tab:eventrate}, the last third and the last rows list the estimates of the corresponding lensing rates for DECIGO and BBO by adopting a conservative threshold of $\delta\varrho_{\rm th}=3$. Apparently, under this more stringent threshold the estimated lensed GW event detection rates decrease by a factor of several to hundred compared with those under the threshold of $\delta \varrho_{\rm th}=1$. However, it is still promising for BBO to detect some lensed systems and thus possible to distinguish different DM models.

Note here the estimation for the expected lensing rate is also affected by other factors, such as the uncertainty in the sBBH merger rate, the choice of the threshold for determining the diffraction cross section, etc. For example, if there are many more sBBH mergers at high redshift than that predicted by the simple model adopted in this paper, the lensing rate can be higher than those listed in Table~\ref{tab:eventrate}. The phase effect is ignored for identifying the lensing events via $\delta\varrho$ in this paper. For a given diffractive lensing system, we expect to obtain a larger $\delta \varrho$ by considering both the amplitude and phase effects and thus may result in a larger detection rate of the lensing events than those obtained in the present paper. Furthermore, we estimate the lensing probability and lensing rate for main haloes but do not consider subhaloes. Low mass subhaloes may also contribute to the lensing rate in this mass range $10^{3}-10^6M_\odot$. In addition, the multiple deflections from weak lensing of all kinds of celestial objects may also contribute unknown noise to signal detection, which needs to be considered in future studies. 

\section{Summary}
\label{sec:sum}

We have systematically investigated the lensed GW signal by mini-halos ($10^3-10^6M_\odot$) in the wave optics regime. We estimate the detection rate of such lensing events by current and future GW detectors assuming different DM models, and we find that the detection rate significantly depends on the DM nature. The reasons for this dependence are the different abundances and the different density profiles of small DM halos resulting from different DM models.

We find that the current and future ground-based GW observatories, such as LIGO, LIGO A+, and ET, are almost unlikely to detect mini-halos (with the NFW profile) via the gravitational lensing of GW within a limited observational period (e.g., less than ten years). CE and GLOC may detect several to tens of GW events diffractively lensed by CDM mini-halos (with the NFW profile) over an observation period of ten years or more. If the mini-halo density profiles has a much steeper (inner) slope than the NFW-like profile, such as the pseudo-Jaffe profile, ET, CE and GLOC are expected to detect a significant number of mini-halos via the diffractive lensing of GWs. With the detection of these lens events, one may be able to distinguish different kinds of DM halo density profiles using the lensed GW signals.

Assuming that the mini-halos resulting from the CDM model follow the NFW density profile, it is expected that the DECIGO/BBO may detect several/hundreds of GW events per year, which are diffractive lensed by mini-halos. Even GLOC can also detect one event per year. For the WDM model ($>10$\,keV) with the same profiles, it is still expected that the DECIGO/BBO may detect about one/several events per year.
As for the SIDM model, the expected lensing rates depend on the choice of the core size for those halos (describing the IC-NFW profile). If the core size is small (e.g., $s=0.1$), the expected lensing rates are similar to those from the CDM model (halos with the NFW profile), especially in the middle-frequency band. If the core size is large (e.g., $s=1$), the expected lensing rates are smaller than those from the CDM model (halos with the NFW profile). However, it is still promising to detect several (tens of) events by the middle-frequency detectors like DECIGO (BBO) if the core size of the SIDM halos is not too large ($s \le1$). These estimates on the detection rates of diffractively lensed GW events by mini-haloes suggest that the DM nature, either cold, warm, or self-interacting, may be significantly constrained via the detection of the diffractively lensed GW events by future GW detectors. In the mean time, the density profiles of mini-haloes as lenses may also be revealed via the lensed GW signals, which would provide further information on the DM nature.

\begin{acknowledgments}		
We thank the anonymous referee for helpful comments and suggestions. We also thank Shun-Sheng Li, Liang Dai, and Shude Mao for stimulating discussions.
This work is partly supported by the National Key Program for Science and Technology Research and Development (Grant No. 2020YFC2201400), the National Natural Science Foundation of China (Grant 11690024, 11873056, 11991052), and the Strategic Priority Program of the Chinese Academy of Sciences (Grant XDB 23040100).
\end{acknowledgments}


\bibliographystyle{apsrev4-2}
\bibliography{refer}

\end{document}